\documentclass[prb,floatfix,twocolumn,showpacs,amsmath,amssymb]{revtex4}
\usepackage{graphicx}
\usepackage{epstopdf}
\usepackage{epsfig}
\usepackage{dcolumn}
\usepackage{bm}
\usepackage{color}

\newcommand{\dagga}{{\phantom{\dagger}}}

\begin{document}

\title{Assessing the orbital selective Mott transition with variational wave functions}
\author{Luca F. Tocchio, Federico Arrigoni, Sandro Sorella, and Federico Becca}
\address{Democritos National Simulation Center, Istituto Officina dei Materiali del CNR, 
and SISSA-International School for Advanced Studies, Via Bonomea 265, I-34136 Trieste, Italy}

\date{\today} 

\begin{abstract}
We study the Mott metal-insulator transition in the two-band Hubbard model with different hopping 
amplitudes $t_1$ and $t_2$ for the two orbitals on the two-dimensional square lattice by using 
{\it non-magnetic} variational wave functions, similarly to what has been considered in the limit
of infinite dimensions by dynamical mean-field theory. We work out the phase diagram at half filling 
(i.e., two electrons per site) as a function of $R=t_2/t_1$ and the on-site Coulomb repulsion $U$, 
for two values of the Hund's coupling $J=0$ and $J/U=0.1$. Our results are in good agreement with 
previous dynamical mean-field theory calculations, demonstrating that the non-magnetic phase diagram 
is only slightly modified from infinite to two spatial dimensions. Three phases are present: 
a metallic one, for small values of $U$, where both orbitals are itinerant; a Mott insulator, for 
large values of $U$, where both orbitals are localized because of the Coulomb repulsion; and the 
so-called orbital-selective Mott insulator (OSMI), for small values of $R$ and intermediate $U$'s, 
where one orbital is localized while the other one is still itinerant. The effect of the Hund's 
coupling is two-fold: on one side, it favors the full Mott phase over the OSMI; on the other side, 
it stabilizes the OSMI at larger values of $R$. 
\end{abstract}

\pacs{71.27.+a, 71.10.Fd, 71.30.+h, 75.25.Dk}

\maketitle

\section{Introduction}\label{sec:intro}

The single-band Hubbard model represents the simplest example where the competition between 
kinetic energy and Coulomb repulsion gives rise to a complex phase diagram, which is still 
representing a formidable problem to be solved in the theory of strongly-correlated systems. 
In many respects, the single-band Hubbard model can be considered to describe materials with 
partially occupied $d$ or $f$ shells. Although this approximation may capture some important aspects 
of these systems, like for example the Mott metal-insulator transition (MIT) and, possibly, the 
existence of $d$-wave superconductivity when doping a Mott insulator, retaining a single band is 
not always sufficient to correctly capture the low-energy properties of materials. Indeed, there 
are many examples in which orbital fluctuations are important and give rise to new physical 
phenomena.~\cite{imada1998,tokura2000} They include cases where the Coulomb exchange, which 
generates the Hund's rules, the existence of crystal fields or Jahn-Teller effects, and bandwidth 
differences among the orbitals produce appreciable effects at low temperatures.

Based on these premises, there is a compelling need to go beyond the single-band Hubbard model in order 
to clarify the role of orbital degeneracy, inter-orbital Coulomb interaction, and Hund's coupling. 
In particular, the MIT in multi-orbital systems involves other energy scales, besides the on-site 
Coulomb repulsion $U$ and the electron bandwidth. In order to highlight the effect of different terms, 
many studies have been performed in the {\it non-magnetic} sector, namely ``neglecting'' any possible 
magnetic long-range order. The aim of this choice is to capture the physics that is driven solely by 
electronic correlation and that can be released when magnetic order is destroyed by the presence of 
frustration (e.g., competing super-exchange couplings). Addressing the properties in the non-magnetic 
sector can be easily considered by different approaches, such as dynamical mean-field theory 
(DMFT)~\cite{georges1996} or slave-boson (SB)~\cite{kotliar1986} techniques, but also within variational 
Monte Carlo (VMC).~\cite{yokoyama1987} For example, whenever magnetic phases are not taken into account, 
DMFT and SB calculations have suggested that the Hund's coupling $J$ has a different effect for different 
filling factors: at half filling, it reduces the value $U_{\rm MIT}$ above which the Mott state is 
stabilized, while, for all the other (integer) fillings, the presence of a finite $J$ increases 
$U_{\rm MIT}$.~\cite{demedici2011a,demedici2011b} Another aspect that induces interesting variations 
in the description of the MIT is the existence of different bandwidths for degenerate 
orbitals.~\cite{liebsch2003} In the past, this issue has been deeply investigated on two-band models on 
the square lattice with hoppings $t_1$ and $t_2$, both intra- and inter-orbital Coulomb repulsions, and 
possibly also the Hund's exchange terms, by means of 
DMFT~\cite{koga2004,ferrero2005,demedici2005,arita2005,liebsch2005,knecht2005,inaba2006} and SB or 
slave-particle approaches.~\cite{demedici2005,ruegg2005} In this case, whenever the ratio $R=t_2/t_1$ is 
sufficiently small (assuming $R \le 1$), the two orbitals have distinct MITs by increasing the Coulomb 
repulsion, which implies the existence of an intermediate phase where one orbital is insulating and 
the other one is metallic; this phase has been named orbital-selective Mott insulator (OSMI). Instead, 
when the two orbitals have comparable hopping amplitudes, i.e., for $R$ larger than a critical value, 
a single MIT is present, where both orbitals undergo a simultaneous transition. Recently, it has been 
proposed that an OSMI can be stabilized also when the orbitals have the same bandwidth, provided they 
have different band dispersions.~\cite{zhang2012} The presence of a crystal-field splitting in the 
Hamiltonian is also responsible for the appearance of an OSMI.~\cite{werner2007,demedici2009,song2009}

The possibility of a phase where some $d$ orbitals give rise to delocalized bands while some others 
remain localized has been discussed in connection with Ca$_{2-x}$Sr$_x$RuO$_4$, to explain the coexistence 
of spin-$1/2$ moments and metallicity at $x=0.5$.~\cite{anisimov2002,nakatsuji2003,yao2013} A partial 
localization of $f$ electrons in some Uranium-based heavy-fermion compounds has been also proposed to 
explain the observed Haas-van Alphen frequencies in UPt$_3$.~\cite{zwicknagl2002} In this context, 
hopping anisotropies driven by intra-atomic correlations have been proposed as the driving mechanism 
for partial localization.~\cite{efremov2004} Moreover, the orbital-selective Mott transition is 
conceptually equivalent to the Kondo breakdown in heavy-fermion systems,~\cite{schroder1998} where the 
localized $f$ electrons suddenly stop to hybridize with the conducting $c$ electrons and no longer 
contribute to the Fermi volume (which is determined by $c$ electrons only).~\cite{vojta2010} 
In this respect, a sign-problem-free model with one itinerant and one fully localized band has been 
studied by Determinant Monte Carlo.~\cite{bouadim2009}

As mentioned, the issue of MITs in multi-orbital models with different hopping amplitudes has been 
investigated mainly by using DMFT, which is exact in infinite dimensions, and SB, which is a simple 
mean-field approximation; by contrast, very few attempts have been done with correlated methods that 
work in finite spatial dimensions.~\cite{koga2006,takenaka2012} In this paper, we examine the phase 
diagram of the Hubbard model in two dimensions, with two degenerate orbitals and $R \le 1$, by using 
correlated variational wave functions that are straightforward generalizations of the Jastrow-Slater 
states that have been widely used to study the single-band Hubbard model in the recent 
past.~\cite{capello2005,capello2006} In particular, the Jastrow factor is considered on top of an 
uncorrelated state, in order to correctly describe the effect of electron-electron interaction. 
Here, the uncorrelated determinant can be factorized into two terms for the different orbitals; 
the crucial ingredient is the inter-orbital Jastrow factor that couples densities on different orbitals 
and allows us a reliable determination of the various phases. 

The outcomes of our variational approach are in good agreement with the ones that have been obtained by 
DMFT.~\cite{ferrero2005,demedici2005,inaba2006} This fact suggests that the (metastable) non-magnetic phase
diagram of the model does not change much from infinite to two dimensions. It is also remarkable that relatively 
simple variational wave functions are able to capture most of the important physical properties also in cases 
where more than one orbital is involved, making it possible to use a similar technique also for other (more 
complicated) multi-orbital systems.

The paper is organized as follows: in Sec.~\ref{sec:model}, we introduce the two-band Hubbard model and the 
variational wave functions that are used to study it; in Sec.~\ref{sec:results}, we present the numerical
results obtained by using variational Monte Carlo for $J=0$ and $J/U=0.1$; finally, 
in Sec.~\ref{sec:conc} we draw our conclusions.
 
\section{Model and method}\label{sec:model}

We consider the two-band Hubbard model defined by:
\begin{equation}\label{eq:hamtot}
{\cal H} = {\cal H}_{\textrm{kin}} + {\cal H}_{\textrm{int}},
\end{equation}
where the kinetic term ${\cal H}_{\textrm{kin}}$ describes hopping processes of electrons within the two 
orbitals:
\begin{equation}\label{eq:hamkin}
{\cal H}_{\textrm{kin}} = - \sum_{\langle i,j\rangle,\alpha,\sigma} t_{\alpha}
c^\dagger_{i,\alpha,\sigma} c^\dagga_{j,\alpha,\sigma} + \rm{h.c.},
\end{equation}
where $c^\dagger_{i,\alpha,\sigma}$ ($c^\dagga_{i,\alpha,\sigma}$) creates (destroys) an electron with 
spin $\sigma$ on site $i$ and orbital $\alpha=1,2$ and $t_{\alpha}$ is the nearest-neighbor hopping amplitude 
with orbital index $\alpha$. We define $R=t_2/t_1$ as the ratio between the two hopping parameters and, 
without loss of generality, we focus on the case with $R\le 1$. In the following, we also fix $t_1=1$.
We would like to stress the fact that the kinetic term is diagonal in the orbital index and, therefore, there 
is no a direct hybridization between different orbitals.

The interaction term includes different contributions:
\begin{eqnarray}
{\cal H}_{\textrm{int}} &=& U \sum_{i,\alpha} n_{i,\alpha,\uparrow}n_{i,\alpha,\downarrow}
                        + U^\prime \sum_{i,\sigma,\sigma^\prime} n_{i,1,\sigma}n_{i,2,\sigma^\prime} \nonumber \\
                        &-& J \sum_{i,\sigma,\sigma^\prime} c^\dagger_{i,1,\sigma} c^\dagga_{i,1,\sigma^\prime}
                                                            c^\dagger_{i,2,\sigma^\prime} c^\dagga_{i,2,\sigma} \nonumber \\
                        &-& J^\prime \sum_{i} (c^\dagger_{i,1,\uparrow} c^\dagger_{i,1,\downarrow}
                                               c^\dagga_{i,2,\uparrow} c^\dagga_{i,2,\downarrow} + \rm{h.c.}),
\label{eq:hamint}
\end{eqnarray}
where $n_{i,\alpha,\sigma}=c^\dagger_{i,\alpha,\sigma} c^\dagga_{i,\alpha,\sigma}$ is the electronic density 
per spin on site $i$ and orbital $\alpha$. These four terms represent the intra-orbital interaction $U$, 
the inter-orbital interaction $U^\prime$, the Hund's coupling $J$, and the pair hopping $J^\prime$. 

In order to study the occurrence of the OSMI, we focus on the half-filled case, i.e., two electrons per site.
There are $6$ atomic states with $2$ particles per site. In particular, the states that diagonalize the
single-site Hamiltonian are one (three-fold degenerate) triplet with energy $E=U^\prime-J$:
\begin{eqnarray}
|1\rangle &=& c^\dagger_{i,1,\uparrow} c^\dagger_{i,2,\uparrow} |0\rangle, \label{eq:triplet1} \\
|2\rangle &=& c^\dagger_{i,1,\downarrow} c^\dagger_{i,2,\downarrow} |0\rangle, \label{eq:triplet2} \\
|3\rangle &=& \frac{1}{\sqrt{2}} \left (c^\dagger_{i,1,\uparrow} c^\dagger_{i,2,\downarrow} -
                                        c^\dagger_{i,2,\uparrow} c^\dagger_{i,1,\downarrow} \right ) |0\rangle, \label{eq:triplet3}
\end{eqnarray}
one singlet with electrons on different orbitals and energy $E=U^\prime+J$:
\begin{equation}
|4\rangle = \frac{1}{\sqrt{2}} \left (c^\dagger_{i,1,\uparrow} c^\dagger_{i,2,\downarrow} +
                                      c^\dagger_{i,2,\uparrow} c^\dagger_{i,1,\downarrow} \right ) |0\rangle, \label{eq:singlet1}
\end{equation}
and, finally, two singlets with electrons on the same orbital and energies $E=U \pm J^\prime$:
\begin{eqnarray}
|5\rangle &=& \frac{1}{\sqrt{2}} \left (c^\dagger_{i,1,\uparrow} c^\dagger_{i,1,\downarrow} -
                                        c^\dagger_{i,2,\uparrow} c^\dagger_{i,2,\downarrow} \right ) |0\rangle, \label{eq:singlet2} \\
|6\rangle &=& \frac{1}{\sqrt{2}} \left (c^\dagger_{i,1,\uparrow} c^\dagger_{i,1,\downarrow} +
                                        c^\dagger_{i,2,\uparrow} c^\dagger_{i,2,\downarrow} \right ) |0\rangle. \label{eq:singlet3}
\end{eqnarray}

According to the rotational symmetry of degenerate orbitals, we set $U^\prime=U-2J$ and 
$J^\prime=J$.~\cite{kanamori1963,castellani1978} In this case, for $J=0$ the ground state of the single 
site is six-fold degenerate, with $E=U$; in the most general case with $J>0$ instead, the ground state is the 
triplet with $E=U-3J$, separated by a doubly-degenerate singlet (i.e., $|4\rangle$ and $|5\rangle$) with 
$E=U-J$; finally the singlet $|6\rangle$ has the highest energy $E=U+J$.

Our numerical results are based on the definition of variational wave functions that approximate the ground-state 
properties beyond perturbative approaches. We consider non-magnetic states as described by the Jastrow-Slater wave 
function that extends the original formulation by Gutzwiller:~\cite{yokoyama1987,gutzwiller1963}
\begin{equation}\label{eq:wavefunction}
|\Psi\rangle={\cal J}|\Phi_0\rangle,  
\end{equation}
where $|\Phi_0\rangle$ is an uncorrelated state that corresponds to the ground state of a BCS Hamiltonian:
\begin{eqnarray}
&& {\cal H}_{\rm BCS } = \sum_{k,\alpha,\sigma} \xi_k^{\alpha} c^\dagger_{k,\alpha,\sigma} c^\dagga_{k,\alpha,\sigma} \nonumber \\
&&                  + \sum_{i,\sigma} \tilde{t}_{\perp} \left (c^\dagger_{i,1,\sigma} c^\dagga_{i,2,\sigma} 
                    + c^\dagger_{i,2,\sigma} c^\dagga_{i,1,\sigma} \right ) \nonumber \\
&&                  + \sum_{k,\alpha} \Delta_k^{\alpha} \left (c^\dagger_{k,\alpha,\uparrow} c^\dagger_{-k,\alpha,\downarrow} 
                    + c^\dagga_{-k,\alpha,\downarrow} c^\dagga_{k,\alpha,\uparrow} \right ) \nonumber \\ 
&&                  + \sum_{i} \left [ (\Delta^{s}_{\perp}+\Delta^{t}_{\perp}) c^\dagger_{i,1,\uparrow}c^\dagger_{i,2,\downarrow} 
                    + (\Delta^{s}_{\perp}-\Delta^{t}_{\perp}) c^\dagger_{i,2,\uparrow} c^\dagger_{i,1,\downarrow} \right . \nonumber \\
&&                  + \left . (\Delta^{s}_{\perp}+\Delta^{t}_{\perp}) c^\dagga_{i,2,\downarrow}c^\dagga_{i,1,\uparrow}  
                    + (\Delta^{s}_{\perp}-\Delta^{t}_{\perp}) c^\dagga_{i,1,\downarrow} c^\dagga_{i,2,\uparrow} \right ].
\label{eq:hambcs}
\end{eqnarray}
The free intra-orbital dispersions $\xi_k^{\alpha}$ are given by:
\begin{equation}\label{eq:xi}
\xi_k^{\alpha} = -2\tilde{t}_{\alpha}[\cos k_x + \cos k_y] -\mu_{\alpha},
\end{equation}
$\tilde{t}_2$, $\mu_1$ ,and $\mu_2$ being variational parameters that are optimized to minimize the variational
energy (while $\tilde{t}_1=1$ sets the energy scale of the BCS Hamiltonian). Even though we allow for the presence of 
a further inter-orbital hopping $\tilde{t}_{\perp}$ in the BCS Hamiltonian~(\ref{eq:hambcs}), we find that the optimal 
variational states are obtained taking $\tilde{t}_{\perp}=0$. The best intra-orbital pairing terms have $d$-wave 
symmetry:
\begin{equation}\label{eq:delta}
\Delta_k^{\alpha}=2\Delta^{\alpha}[\cos k_x - \cos k_y],
\end{equation}
with $\Delta_k^{\alpha}$ being further variational parameters, in analogy with the one-band Hubbard model for the 
square lattice.~\cite{gros1988,zhang1988} In addition, inter-orbital terms are considered, either with singlet (i.e., 
$\Delta^{s}_{\perp}$) or with triplet (i.e., $\Delta^{t}_{\perp}$) symmetries. As a result, we find that the 
inter-orbital singlet component is never relevant in the phase diagram, while the triplet component is necessary 
to correctly describe the full Mott phase in the presence of a finite Hund's coupling $J$, as pointed out also by 
the Gutzwiller approximation.~\cite{zegrodnik2014}

The effects of correlations are introduced by means of the so-called Jastrow factor ${\cal J}$:
\begin{equation}\label{eq:jastrow}
{\cal J} = \exp \left ( -\frac{1}{2} \sum_{i,j,\alpha,\beta} 
v^{\alpha,\beta}_{i,j} n_{i,\alpha} n_{j,\beta} \right ),
\end{equation}
where $n_{i,\alpha}= \sum_{\sigma} n_{i,\alpha,\sigma}$ is the electron density on site $i$ and orbital 
$\alpha$; $v^{\alpha,\beta}_{i,j} = v^{\beta,\alpha}_{i,j}$ (that include also the local Gutzwiller term for 
$\alpha=\beta$ and $i=j$) are pseudopotentials that are optimized for every independent distance 
$|{\bf R}_i-{\bf R}_j|$. The Jastrow factor has been shown to be crucial in describing a Mott insulating 
state within the single-band Hubbard model, with $v_q \propto 1/q^2$ ($v_q$ being the Fourier transform of 
$v_{i,j}$) in the insulating region, while $v_q \propto 1/q$ is found in the metallic/superconducting 
phase.~\cite{capello2005,capello2006} We remark that the Jastrow factor embodies a crucial long-range 
attraction between doubly occupied and empty sites, keeping them bounded in the Mott 
phase.~\cite{capello2005,capello2007,yokoyama2011,miyagawa2011} Otherwise, a non-magnetic Mott insulator 
cannot be obtained by using only a local (i.e., on-site) Gutzwiller term.~\cite{yokoyama1987} Similar results 
have been obtained in the bilayer Hubbard model.~\cite{rueger2014} Also in this paper on the two-band Hubbard 
model, we show that the presence of a Jastrow factor is necessary to obtain a non-trivial phase diagram with 
metallic and Mott phases, as well as the OSMI. We remark that, while an insulating state can be obtained just 
by applying the Jastrow factor on top of a Fermi gas, the pairing terms of the BCS Hamiltonian are particularly 
important to have a correct description of Mott insulators, where localized electrons are paired together to 
form the RVB state, as proposed originally by Anderson.~\cite{anderson1987}

We mention that the Jastrow-Slater wave functions of Eq.~(\ref{eq:wavefunction}) may be improved by considering 
the so-called backflow correlations,~\cite{tocchio2008,tocchio2011} which are particularly important at half 
filling and for small hole dopings; however, the physical content of the original variational states is not 
modified by the inclusion of backflow terms. Therefore, in the following we will not consider these corrections, 
which are relatively computationally expensive.

The wave function obtained from applying the Jastrow factor to the ground state of the BCS Hamiltonian of 
Eq.~(\ref{eq:hambcs}) does not describe phases with magnetic long-range order; also phases with orbital order 
cannot be captured. In this sense, as discussed in the introduction, our variational states are suitable to
approach the paramagnetic Mott transition, driven solely by electronic correlations. We would like to mention
the fact that orbital order may be obtained whenever, in the BCS Hamiltonian~(\ref{eq:hambcs}), the $d$-wave 
intra-orbital pairing is replaced by an on-site $s$-wave one:
\begin{equation}\label{eq:s-wave}
  \sum_{k} \left[ \Delta_1 c^\dagger_{k,1,\uparrow} c^\dagger_{-k,1,\downarrow} +
                  \Delta_2 c^\dagger_{k,2,\uparrow} c^\dagger_{-k,2,\downarrow} + \rm{h.c.} \right].
\end{equation}
Indeed, this pairing term gives a sizable energy gain for $J=0$, due to the appearance of a staggered 
orbital order, where the orbital $1$ is (almost) doubly occupied and the orbital $2$ is (almost) empty on one 
sublattice and vice-versa for the other sublattice. Remarkably, although the BCS Hamiltonian of Eq.~(\ref{eq:s-wave}) is translationally
invariant (implying a translationally invariant $|\Phi_0\rangle$), density-density correlations computed with the 
wave function $|\Psi\rangle$ of Eq.~(\ref{eq:wavefunction}) clearly show long-range orbital order. The fact that (correlated) 
translationally invariant wave functions may show long-range order has been already discussed in one-band models, 
where dimerization~\cite{capello2005} or charge order~\cite{tocchio2014} can be obtained, and has been investigated 
in detail in Ref.~\cite{kaneko2015}.

\begin{figure}
\centering
\includegraphics[width=1.0\columnwidth]{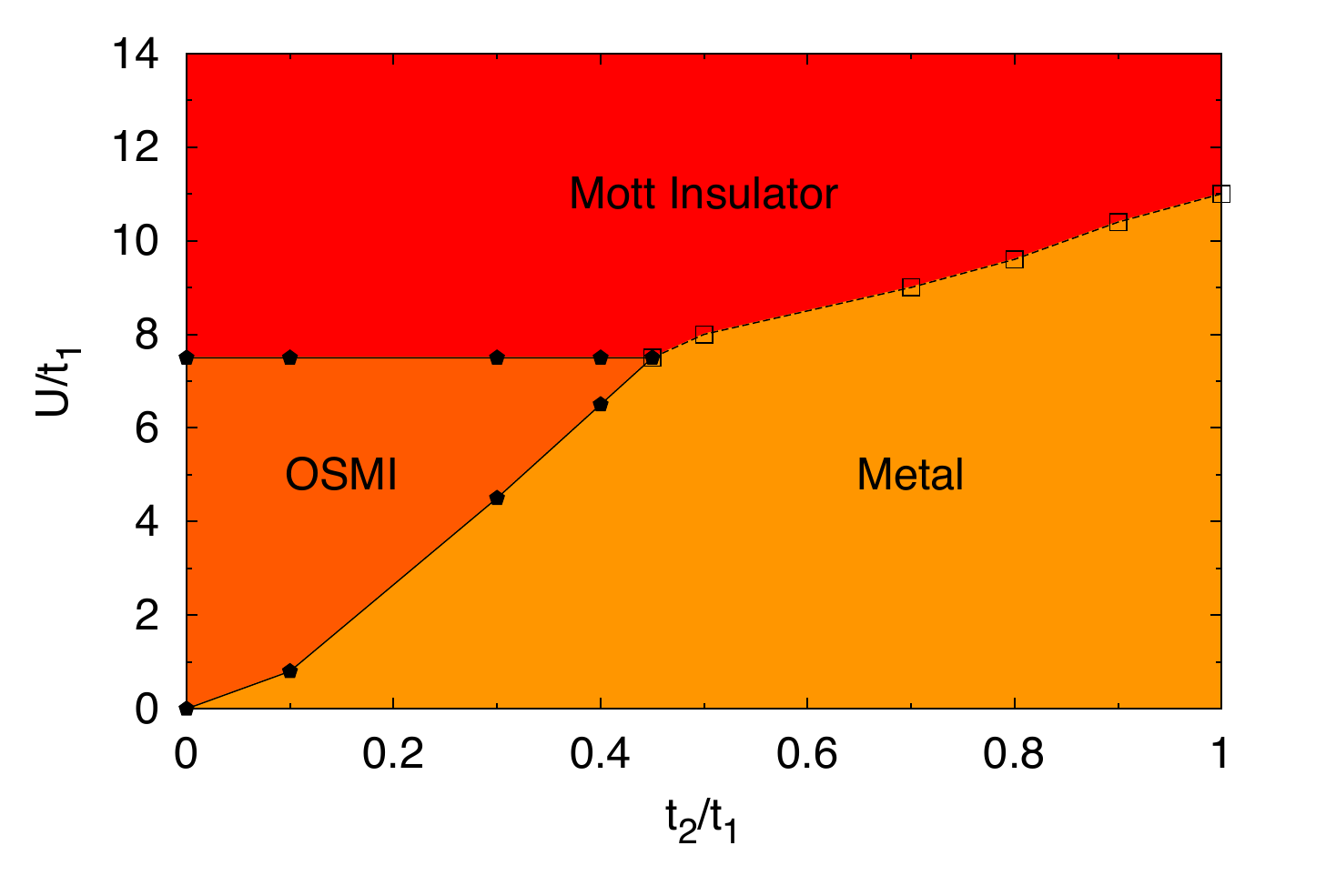}
\caption{\label{fig:pdj0}
(Color online) Non-magnetic phase diagram of the two-band Hubbard model with $J=0$. Three regions can be 
identified as a function of $R$ and $U/t_1$: a metal (where both orbitals are metallic), a full Mott insulator 
(where both orbitals are insulating), and the orbital-selective Mott insulator (where the orbital with the 
smallest bandwidth is insulating while the one with the largest bandwidth is metallic). Continuous lines 
denote second-order transitions, while the dashed line denotes a first-order transition.} 
\end{figure}

\begin{figure}
\centering
\includegraphics[width=0.8\columnwidth]{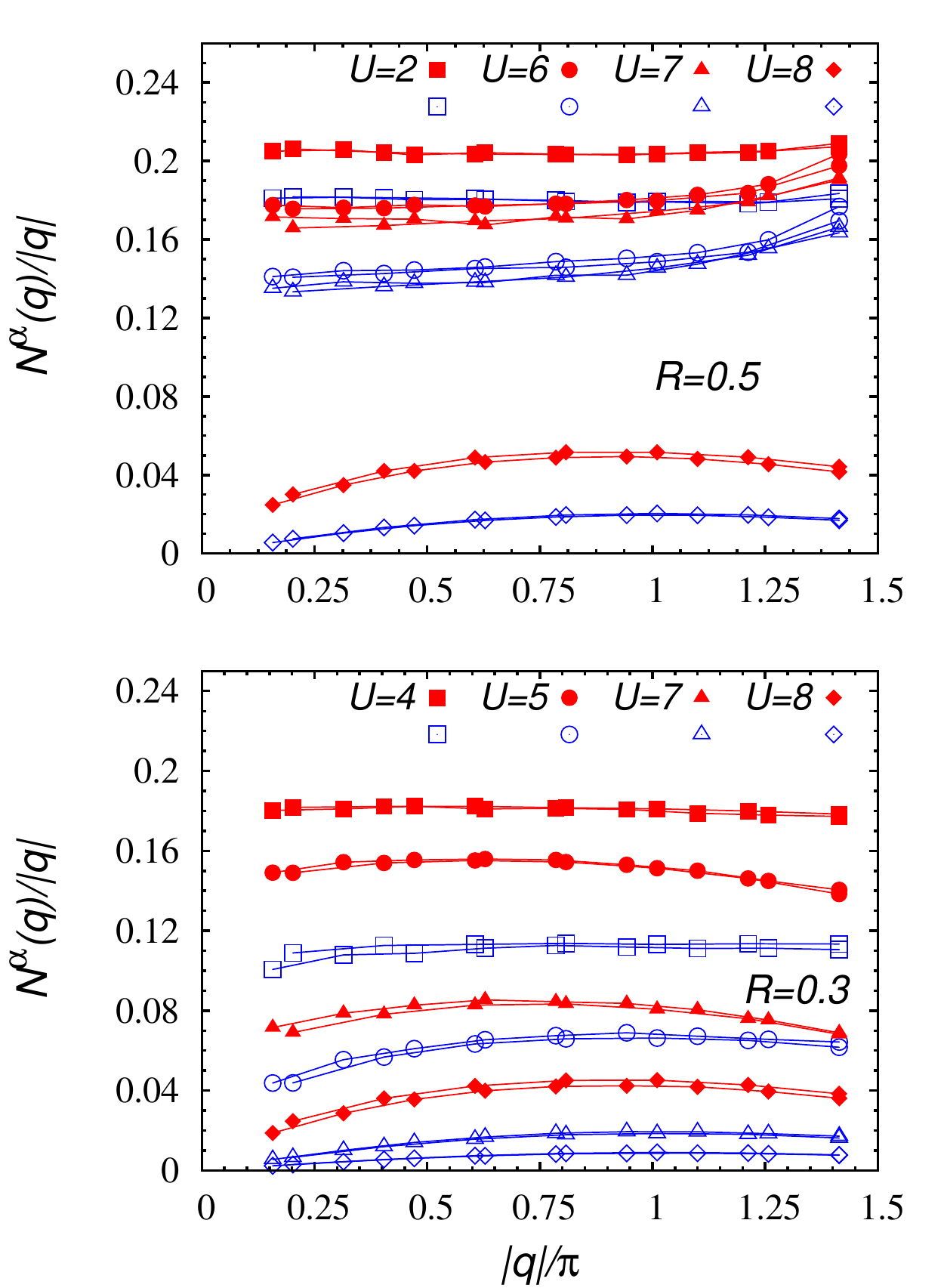}
\caption{\label{fig:Nqj0}
(Color online) Intra-orbital density-density structure factor $N^{\alpha}(q)$ divided by $|q|$ as a function of 
$|q|/\pi$ along the $\Gamma{-}X$ line, where $\Gamma=(0,0)$ and $X=(\pi,\pi)$  in the first Brillouin zone. 
Data are shown for different values of $U/t_1$ at $R=0.3$ and at $R=0.5$ for $J=0$. Red full symbols refer to 
the case $\alpha=1$, while empty blue symbols refer to $\alpha=2$. Results are shown for $L=98$ and $162$. 
Statistical error bars are smaller than the symbol size.}
\end{figure}

Finally, we would like to emphasize the advantages and disadvantages of the variational Monte Carlo method.
The main advantage is that correlated states may be considered and treated beyond any perturbative approach 
and without any approximation (e.g., without the Gutzwiller approximation~\cite{brinkman1970,vollhardt1984}). 
However, in order to compute expectation values over variational states, a Monte Carlo sampling is necessary, 
thus leading to statistical errors. The energy computed with variational Monte Carlo gives an upper bound to 
the exact value, thus providing a criterion to judge the quality of the variational states. Moreover, it is 
possible to assess quite large clusters, with all relevant spatial symmetries (translations, rotations, and 
reflections) preserved. By contrast, it is difficult to quantify the systematic errors, which are introduced 
by the choice of the trial state.

\section{Results}\label{sec:results}

In this section, we present the variational results obtained by using the Jastrow-Slater wave function of
Eq.~(\ref{eq:wavefunction}). We study the model on two-dimensional square lattices with $L$ sites and take 
$45$-degree tilted clusters with $L=2 l^2$ sites, $l$ being an odd integer. First, we consider the case with 
$J=0$, then we study the effect of a small Hund's coupling, i.e., $J/U=0.1$.

\subsection{The case with $J=0$}\label{sec:J=0}

Let us start by pointing out that, if no inter-orbital coupling is present in the Hamiltonian of 
Eq.~(\ref{eq:hamint}) (i.e., $U^\prime=0$ and $J=J^\prime=0$), the OSMI would take place in a quite large 
region of the phase diagram. Indeed, the full Hamiltonian~(\ref{eq:hamtot}) would decouple into two 
single-band Hubbard models, with the same Coulomb repulsion $U$ but different hopping amplitudes, e.g., 
$R < 1$. In the non-magnetic sector, the two orbitals would have distinct MITs, because 
$R \ne 1$. The phase diagram, in the $(R,U/t_1)$ plane would be very simple: i) a Mott phase for 
$U>U_{\rm MIT}$, where $U_{\rm MIT}$ is the critical value for the single-band model; ii) a metallic phase 
for $U<U_{\rm MIT} \, R$; and iii) an OSMI for $U_{\rm MIT} \, R<U<U_{\rm MIT}$. These (trivial) results 
are obtained within the variational wave function~(\ref{eq:wavefunction}) by imposing a vanishing 
inter-orbital Jastrow factor in Eq.~(\ref{eq:jastrow}), i.e., $v^{\alpha,\beta}_{i,j}=0$ for 
$\alpha \ne \beta$. In this case, $U_{\rm MIT}/t_1=7.5 \pm 0.5$.~\cite{noteUc}

The results are substantially modified in the presence of the inter-orbital coupling $U^\prime=U$ ($J=0$), 
which favors the metallic phase over a much larger region. Within the variational approach, this effect is 
captured by allowing an inter-orbital Jastrow factor $v^{\alpha,\beta}_{i,j}$ with $\alpha \ne \beta$ in 
Eq.~(\ref{eq:jastrow}). Our results, obtained from calculations on $98$ and $162$ sites (with two orbitals per 
site) are summarized in Fig.~\ref{fig:pdj0}, where we report the ground-state phase diagram in the 
$(R,U/t_1)$ plane. We notice that, as long as the value of $R$ is sufficiently small, e.g., $R \lesssim 0.5$, 
the two orbitals stay essentially decoupled, and the OSMI may exist at intermediate Coulomb interactions. 
In addition, the critical $U$ leading to the full Mott phase does not depend upon $R$, as expected, since the 
two orbitals are decoupled. By contrast, for $R \gtrsim 0.5$, the OSMI disappears, given the effective 
hybridization between the two orbitals. Here, the value of $U_{\rm MIT}$, at which the Mott state takes place, 
increases monotonically with $R$. This result is consistent with what has been suggested by a Monte Carlo 
study of the degenerate $M$-band Hubbard model on the square lattice with $R=1$, where $U_{\rm MIT}/W \simeq \sqrt{M}$ 
($W=8t_1$ being the bandwidth).~\cite{gunnarsson1996} For $M=2$ (and $R=1$), a larger value for the MIT, i.e., 
$U_{\rm MIT}/W \simeq 2$, has been instead predicted by the slave-boson approach of Ref.~\cite{klejnberg1998}. 
Our present results for the whole phase diagram are qualitatively similar to what has been obtained within DMFT 
by several authors in the past.~\cite{ferrero2005,demedici2005,inaba2006} Some quantitative discrepancies may 
be observed in the location of the transition to the full Mott insulator, that is located at slightly higher 
values of $U/t_1$ in our approach. Moreover, we find a larger region of OSMI, that extends to $R \approx 0.45$ 
instead of $R \approx 0.2$, as obtained in DMFT.

\begin{figure}
\centering
\includegraphics[width=0.8\columnwidth]{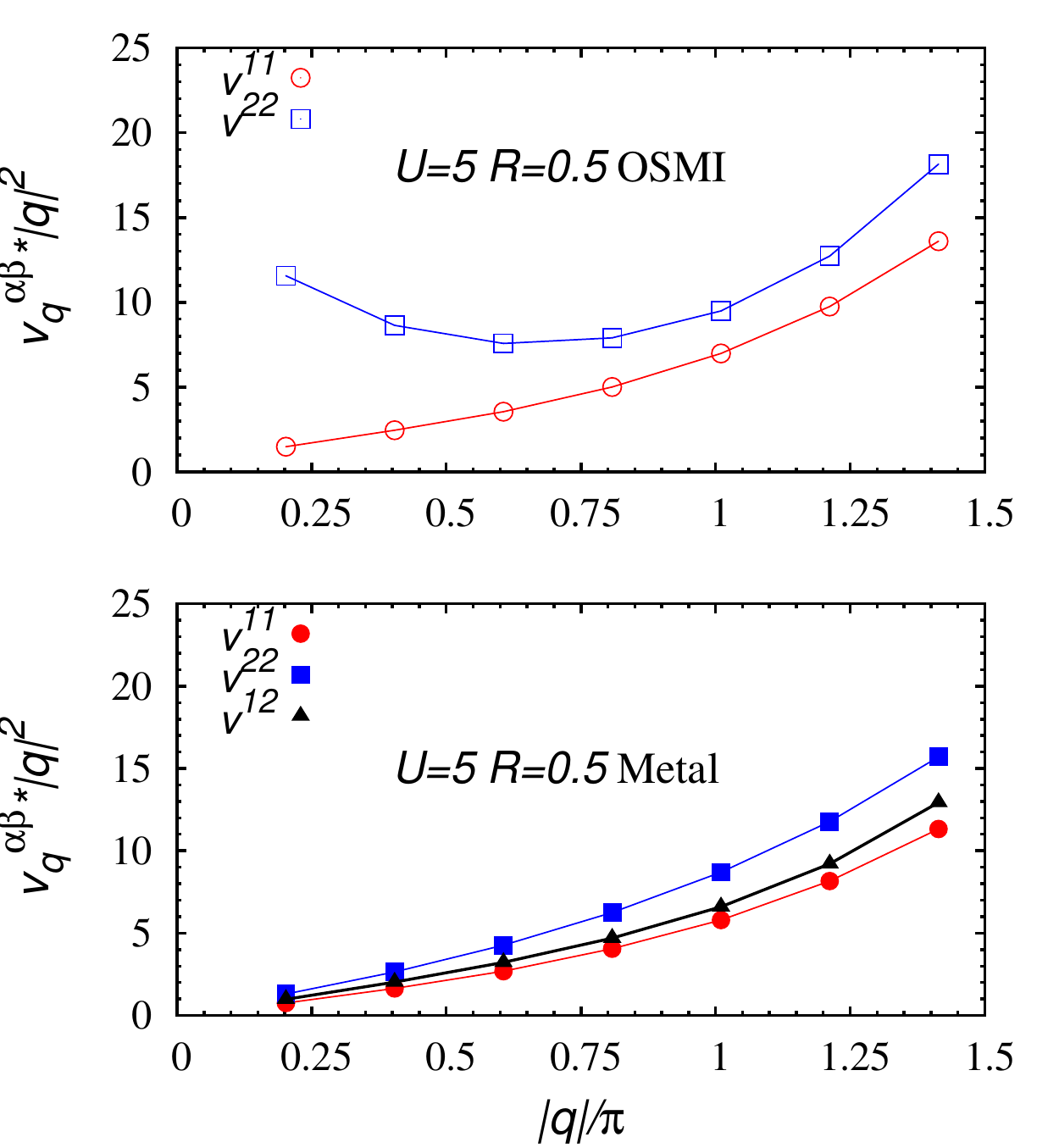}
\caption{\label{fig:Jastrow}
(Color online) The case with $U/t_1=5$ and $R=0.5$, which falls in the OSMI if the two orbitals are not 
interacting, i.e., $U^\prime=0$ and $J=0$, while it falls in the metallic phase for $U^\prime=U$ ($J=0$). 
Upper panel: intra-orbital Jastrow factors $v^{\alpha,\alpha}_q$ multiplied by $|q|^2$, as a function of 
$|q|/\pi$, for the case where a variational state is taken with $v^{1,2}_{i,j}=0$. In this case the 
$\alpha=1$ orbital is metallic (i.e., $v^{1,1}_q \propto 1/|q|$), while the $\alpha=2$ orbital is insulating 
(i.e., $v^{2,2}_q \propto 1/q^2$). Lower panel: Jastrow factors $v^{\alpha,\beta}_q$ multiplied by $|q|^2$, 
as a function of $|q|/\pi$, for the case where all Jastrow terms are optimized together. Here, the presence 
of the inter-orbital Jastrow factor stabilizes a metallic state, with $v^{\alpha,\alpha}_q \propto 1/|q|$.
In both panels, the $q$ points are along the $\Gamma{-}X$ line, with $\Gamma=(0,0)$ and $X=(\pi,\pi)$. 
The results are obtained for $L=98$. Statistical error bars are smaller than the symbol size.} 
\end{figure}

We remark that the OSMI is stable also when including a direct (on-site) inter-orbital hopping 
$\tilde{t}_{\perp}$ in the mean-field Hamiltonian of Eq.~(\ref{eq:hambcs}). Indeed, even if this possibility 
is allowed, the optimal variational state has $\tilde{t}_{\perp}=0$. In addition, we also verified that each 
orbital remains half-filled even if charge transfer processes are allowed within the Monte Carlo moves. 

In practice, the metallic or insulating nature can be determined by looking at the static density-density 
structure factor.~\cite{capello2005,capello2006,tocchio2011} Let us briefly discuss this issue on the 
single-band Hubbard model and then generalize it to the two-band case. Consider:
\begin{equation}
N(q) = \frac{\langle \Psi| n_{-q}n_{q} |\Psi \rangle}{\langle \Psi|\Psi \rangle},
\end{equation} 
where $n_q=1/\sqrt{L}\sum_{r} e^{iqr}n_{r}$ is the Fourier transform of the particle density 
$n_{r}$ (on a single-band model there is only one orbital and we drop the index for that). A metallic 
behavior is characterized by $N(q) \propto |q|$ for $q\to 0$, which implies a vanishing gap for particle-hole 
excitations. On the contrary, $N(q)\propto q^2$ for $q \to 0$ implies a finite gap and an insulating behavior. 
These facts are a consequence of approximating the lowest-energy excitations by the Feynman construction
(the so-called single-mode approximation) as $|\Psi_q \rangle = n_q|\Psi \rangle$.~\cite{feynman1954}

Here, we are considering a model where there is no direct hybridization between the two orbitals in 
the $J=0$ case and where the charge transfer between the orbitals remains negligible also in the presence 
of the pair hopping term in the Hamiltonian. Therefore, since we are interested in the properties of each 
orbital individually, we consider only the intra-orbital correlations:
\begin{equation}\label{eq:nqnq}
N^{\alpha}(q)=\frac{\langle \Psi| n_{\alpha,-q} n_{\alpha,q} |\Psi \rangle}{\langle \Psi|\Psi \rangle}, 
\end{equation}
where now $n_{\alpha,q}$ is the Fourier transform of the particle density on the orbital $\alpha$. The metallic 
or insulating behavior of the two orbitals can be assessed by looking at the small-$q$ behavior of 
$N^{\alpha}(q)$. The metallic (Mott) phase is characterized by $N^{\alpha}(q) \propto |q|$ 
($N^{\alpha}(q) \propto q^2$) for both $\alpha=1$ and $2$, while the OSMI has $N^{1}(q) \propto |q|$ and 
$N^{2}(q) \propto q^2$. 

\begin{figure}
\centering
\includegraphics[width=0.8\columnwidth]{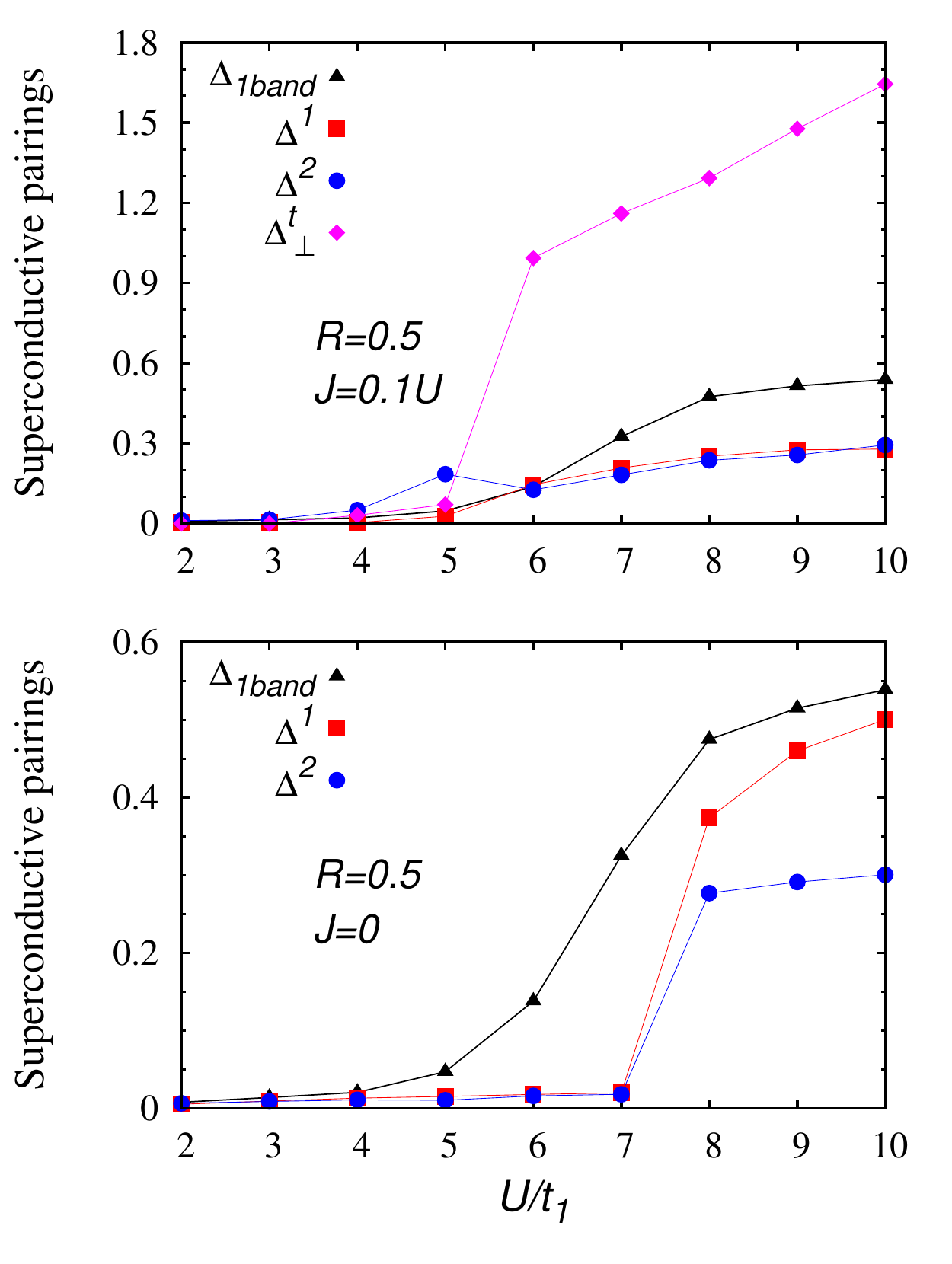}
\caption{\label{fig:smooth}
(Color online) Superconducting pairing fields in the variational state as a function of $U/t_1$ for $R=0.5$ 
with $J/U=0.1$ (upper panel) and $J=0$ (lower panel). $\Delta^{1}$ and $\Delta^{2}$ represent intra-orbital 
$d$-wave pairing in the orbital $1$ and $2$, respectively; $\Delta^t_{\perp}$ represents triplet pairing 
between different orbitals on the same site (shown only for $J/U=0.1$). The $d$-wave pairing field of the 
one-band case $\Delta_{1band}$ is also shown for comparison. Data are presented for a $L=98$ lattice size. 
Statistical error bars are smaller than the symbol size.}
\end{figure}

In Fig.~\ref{fig:Nqj0} data are shown for the two cases $R=0.3$ and $R=0.5$. While in the first case an 
intermediate phase, with one orbital that is metallic and the other one that is insulating, can be identified 
as a function of $U/t_1$, in the second one, a direct transition between a fully metallic phase and a full 
Mott state is observed. Moreover, in the first case with $R=0.3$ the coefficient of the linear term in 
$N^{\alpha}(q)$ goes to zero smoothly when approaching the metal-insulator transition (for both bands). 
This result suggests that both the transition between the metallic and the OSMI phases and the one between 
the OSMI and the Mott state are second order. This statement is further supported by the fact that no fully 
metallic state can be stabilized within the OSMI phase (and no OSMI can be stabilized within the Mott phase), 
as a local minimum. By contrast, for $R=0.5$ the coefficient of the linear term in both $N^{1}(q)$ and 
$N^{2}(q)$ suddenly drops at the transition. In addition, metastable solutions can be found, which indicates
that the metal-insulator transition is first order.

The large metallic region that is observed in the phase diagram can be stabilized by the presence of the 
inter-orbital Jastrow factor. Indeed, a variational wave function without this term would show a much more 
extended region of OSMI, with a significantly higher variational energy. For example, for $U/t_1=5$ and 
$R=0.5$ the wave function without inter-orbital Jastrow factor has an energy that is $0.1t_1$ higher than the 
best metallic solution with all Jastrow factors. In Fig.~\ref{fig:Jastrow}, we report the calculations for 
the Jastrow parameters in these two cases. We show that the presence of an inter-orbital Jastrow factor 
$v^{1,2}_{i,j}$ in the variational state is able to change the small-$q$ behavior of the intra-orbital 
Jastrow factor for the most correlated band, namely $v^{2,2}_q$, thus leading to a metallic behavior in both 
bands. 

\begin{figure}
\centering
\includegraphics[width=1.0\columnwidth]{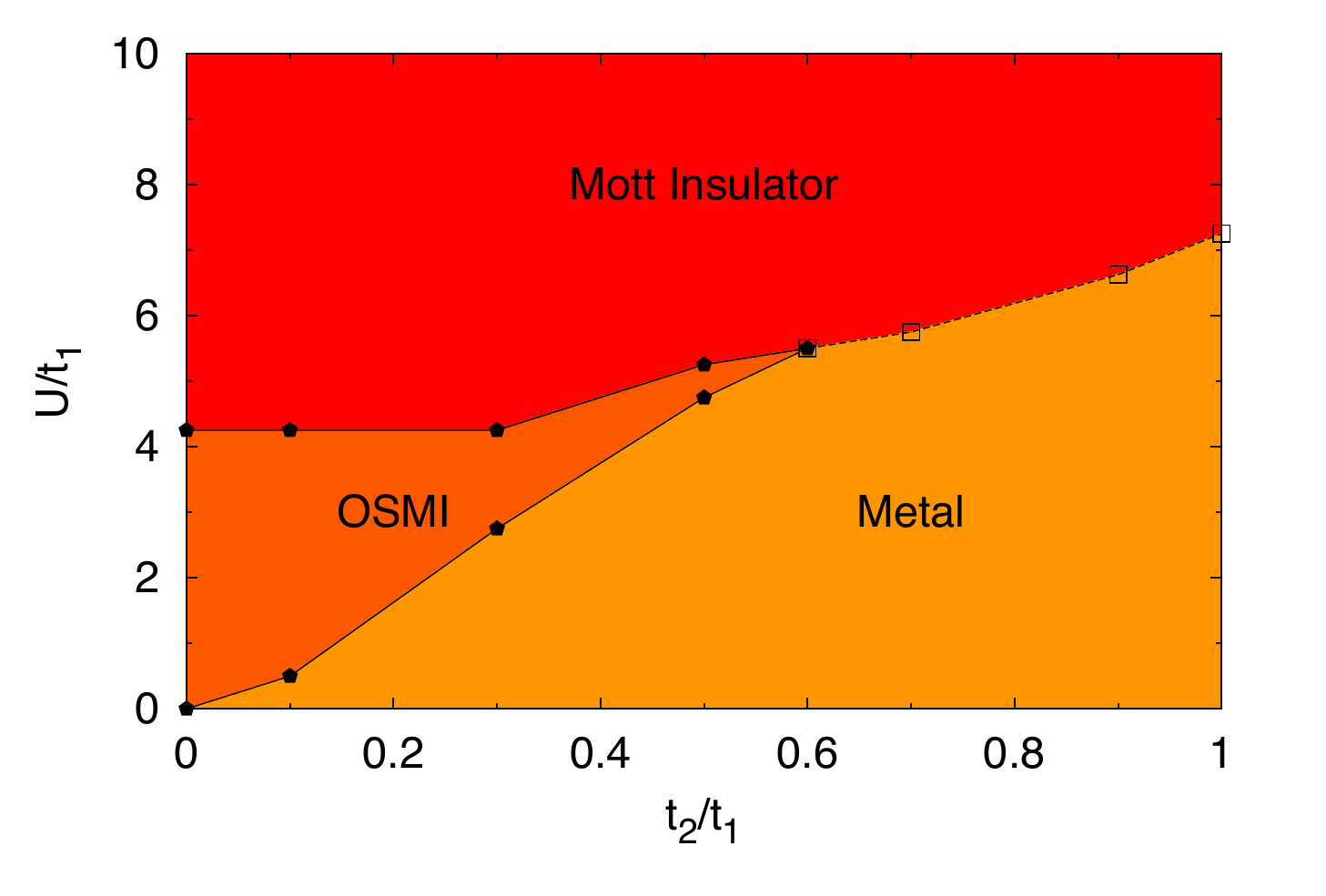}
\caption{\label{fig:pdj1}
(Color online) The same as in Fig.~\ref{fig:pdj0} but with $J/U=0.1$.}
\end{figure}

\begin{figure}
\centering
\includegraphics[width=0.8\columnwidth]{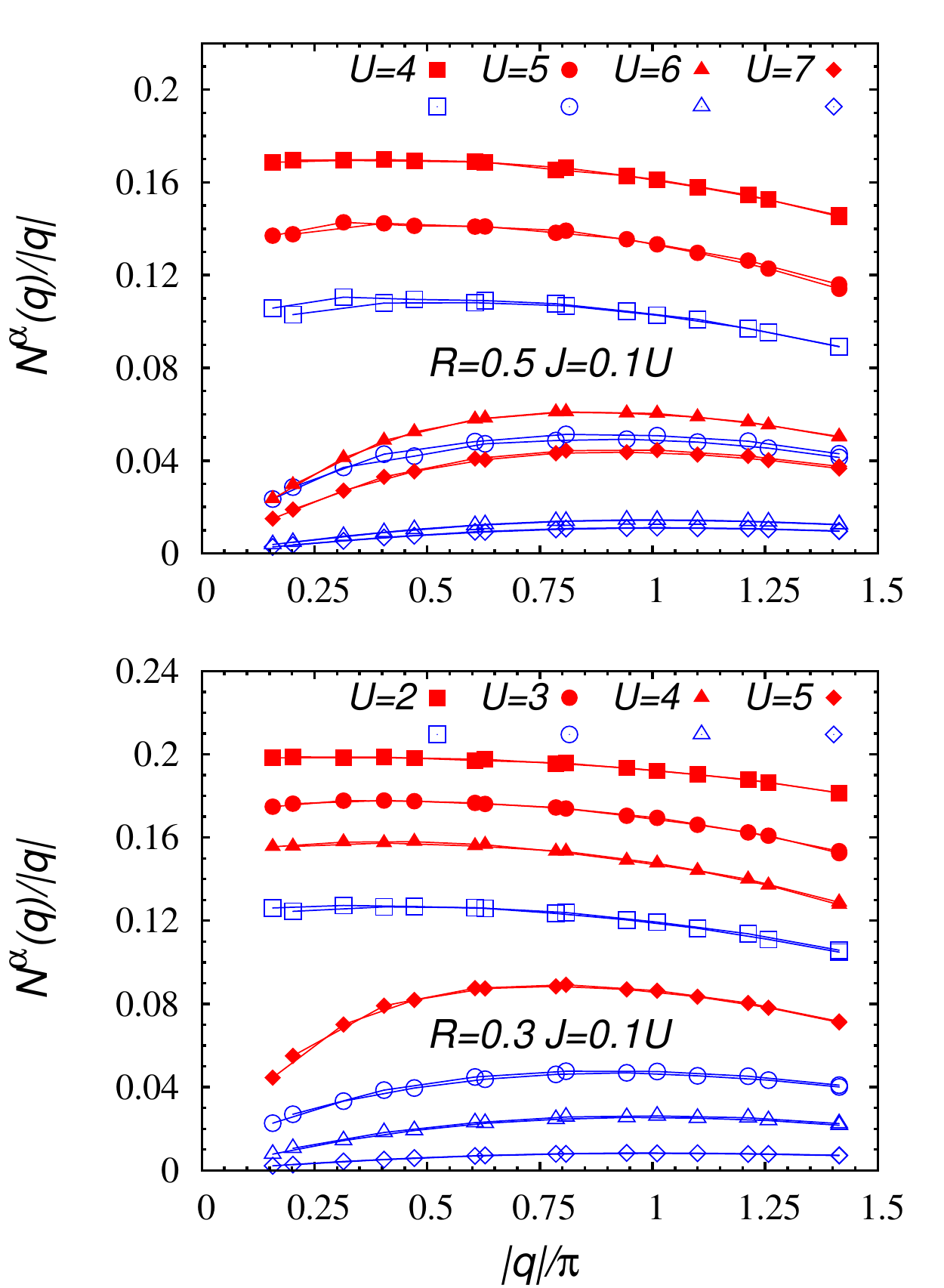}
\caption{\label{fig:Nqj1}
(Color online) The same as in Fig.~\ref{fig:Nqj0} but with $J/U=0.1$.}
\end{figure}

Concerning the mean-field part of the variational state, we show in Fig.~\ref{fig:smooth} (lower panel) 
the behavior of the intra-orbital BCS pairings $\Delta^1$ and $\Delta^2$ of Eq.~(\ref{eq:delta}), as 
a function of $U/t_1$, for a value of $R$ where a direct transition between a metallic and the full Mott 
state occurs. We notice that both the intra-orbital pairings have a jump at the transition, from a vanishingly 
small value in the metallic phase to a finite value when the bands are insulating. This behavior is different 
from what happens in the one-band Hubbard model, in which the BCS pairing increases smoothly with $U$, 
indicating a second-order transition in the thermodynamic limit.

We finally remark that an orbital ordered state can be obtained by allowing for an on-site $s$-wave
pairing field, as described in Eq.~(\ref{eq:s-wave}). Indeed, the presence of this term leads to a 
considerable energy gain and induces long-range orbital order with $n_{i,\alpha} \propto [1+(-1)^{R_i+\alpha}]$. 
The presence of orbital order can be detected by looking at the intra-orbital density-density correlations 
$N^{\alpha}(q)$. Indeed, the characteristic density pattern within each orbital, with doubly-occupied sites 
surrounded by empty sites, is reflected into a divergent peak of $N^{\alpha}(q)$ at the vector $Q=(\pi,\pi)$ 
(not shown). Even if we have evidence that orbital order would occur for any value of $U$ and $R$, a precise 
size scaling of $N^{\alpha}(Q)$ for small values of $U$ would require larger lattice sizes and is out of the 
scope of the present paper (that focuses on the paramagnetic Mott transition).
  
\subsection{The case with $J/U=0.1$}\label{sec:J=01}

Let us now turn to the case with a finite Hund's coupling $J$ and consider the case with $J/U=0.1$. 
The ground-state phase diagram in the $(R,U/t_1)$ plane is reported in Fig.~\ref{fig:pdj1}, to be compared 
with the one for $J=0$ in Fig.~\ref{fig:pdj0}. Moreover, in Fig.~\ref{fig:Nqj1}, we report the results for
the density-density correlations of Eq.~(\ref{eq:nqnq}) for $R=0.3$ and $R=0.5$, to be compared with the 
case for $J=0$ in Fig.~\ref{fig:Nqj0}. Two remarks can be drawn. i) The presence of a finite $J$ term 
favors the full Mott state over both the OSMI and the metallic phase. As a result, the transition line that 
marks the stabilization of the Mott phase shifts down to lower values of $U/t_1$. This outcome can be easily 
understood from the fact that the Mott state, where all electrons are localized, has a large energy gain coming
from the Hund's rule, which favors a spin alignment. ii) A finite $J$ coupling also favors the OSMI with 
respect to the metallic phase (i.e., the OSMI can be stabilized for larger values of $R$, up to $0.6$, with 
respect to the $J=0$ case). Within DMFT, this fact has been explained by a non-vanishing magnetic moment in 
the metallic phase when $J>0$,~\cite{demedici2011b} which may gain energy when coupled together with the one 
present in the insulating orbital. Moreover, we have that the critical $U$ that leads to the Mott phase for 
small $R$ is no longer independent from $R$: here, $J$ directly couples the two orbitals and the transition 
point changes from $U_{\rm MIT}/t_1=4 \pm 0.5$ for $R \approx 0$ to $U_{\rm MIT}/t_1=5.5 \pm 0.5$ for $R=0.6$.
This feature is somehow missing in the DMFT picture where the transition to the Mott phase is almost constant 
at $U_{\rm MIT}/t_1\sim 4$.~\cite{inaba2006}. We also remark that the critical $U$ predicted by our Monte Carlo 
approach for the MIT at $R=1$ is smaller than the slave-boson result, where $U/W \simeq 1.3$.~\cite{klejnberg1998} 

One important aspect is that a remarkable energy gain in the Mott phase is obtained by considering an on-site 
and inter-orbital {\it triplet} pairing $\Delta^{t}_{\perp}$ in the mean-field Hamiltonian~(\ref{eq:hambcs}).
This outcome is natural, given the fact that for $J>0$, the atomic ground state of Eq.~(\ref{eq:hamint}) is 
given by the triplet states of Eqs.~(\ref{eq:triplet1}), (\ref{eq:triplet2}), and~(\ref{eq:triplet3}).

In Fig.~\ref{fig:smooth} (upper panel) we report the BCS pairings as a function of $U/t_1$ at $R=0.5$. Three 
different regimes can be distinguished by increasing the Coulomb repulsion: a metallic phase for $U/t_1 \le 4$ 
where all the pairings are negligible, an OSMI phase at $U/t_1 \approx 5$, where the largest pairing is the 
intra-orbital one on the most correlated band, and the Mott insulator, where in addition to the two $d$-wave 
intra-orbital pairings there is a large triplet pairing between different orbitals on the same site. This latter 
term encodes the ferromagnetic Hund's coupling part of the Hamiltonian. We finally remark that, in contrast to
the $J=0$ case, no orbital order is observed for $J/U=0.1$, since this ordered state would be incompatible 
with the Hund's coupling, which favors triplet states. 

\section{Conclusions}\label{sec:conc}

In this paper, we examined a two-band Hubbard model in the case where the two orbitals have different hopping 
amplitudes, with particular emphasis on the existence of the orbital selective Mott insulator, that emerges 
in the non-magnetic sector. In the recent past, this topic has been widely addressed by mean-field methods, 
including SB approaches and DMFT, which is exact in infinite dimensions. Here, we made use of an alternative 
approach, based on variational wave functions with Jastrow terms, in order to capture long-range spatial 
correlations in two dimensions, thus providing a complementary approach to DMFT. The first outcome of our 
study is that the non-magnetic phase diagram does not qualitatively change when going from infinite to two 
spatial dimensions: we confirm the existence of the OSMI phase already for the $J=0$ case; in addition we
verify that the Hund's coupling is favoring the full Mott phase over the OSMI and the OSMI over the metal. 
The second outcome is more technical and refers to the fact that relatively simple variational wave functions 
are able to capture the important physical properties of multi-band Hubbard models, with different kinds of 
interactions. In particular, we highlighted the role of the inter-band Jastrow factor in properly describing 
the orbital hybridization and the role of the triplet inter-orbital pairing in capturing the effect of the 
Hund's coupling. 

Our variational states can be naturally extended to describe three- or even five-orbital models, which are
suitable to describe electrons in partially occupied $d$ shells. These settings would allow, for example, to 
study transition-metal oxides of $3d$ and $4d$ series (such as Cuprates and Ruthenates), as well as the
iron-based superconductors (Iron Pnictides and Chalcogenides).~\cite{georges2013} Finally, with the further 
inclusion of spin-orbit coupling, a three-band model is also appropriate to study the unconventional physics 
of $5d$ materials, such as Iridates.~\cite{watanabe2013,sato2015}

\acknowledgments

We would like to thank M. Fabrizio, R. Valent\'i, C. Gros, L. de' Medici, and J. Mravlje for useful discussions. 
We acknowledge support from PRIN 2010 2010LLKJBX.

\end{document}